\documentclass[12pt]{article}
\usepackage{amssymb,amsmath,amsthm,amsfonts,latexsym}
\def\C{\mathbb{C}}
\def\I{\mathbb{I}}
\def\N{\mathbb{N}}
\def\R{\mathbb{R}}

\begin{document}
\thispagestyle{empty}
\hfill \today

\vspace{2.5cm}

\begin{center}
\bf{\LARGE Quantum Physics and Signal Processing\\[0.3cm]
in Rigged Hilbert Spaces by means of \\[0.3cm]Special Functions, Lie Algebras and \\[0.3cm]
Fourier and Fourier-like Transforms}\footnote{ Contribution to the 30th International Colloquium on Group Theoretical Methods in Physics, July 14--18, 2014, Gent (Belgium)}
\end{center}

\bigskip\bigskip

\begin{center}
E. Celeghini$^{1,2}$, M.A. del Olmo$^2$ 
\end{center}

\begin{center}
$^1${\sl Dipartimento di Fisica, Universit\`a  di Firenze and
INFN--Sezione di
Firenze \\
I50019 Sesto Fiorentino,  Firenze, Italy}\\
\medskip

$^2${\sl Departamento de F\'{\i}sica Te\'orica and IMUVA, Universidad de
Valladolid, \\
E-47011, Valladolid, Spain.}\\
\medskip

{e-mail: celeghini@fi.infn.it, olmo@fta.uva.es}

\end{center}

\bigskip

\bigskip

\begin{abstract}
Quantum Mechanics and Signal Processing in the line\, $\R$\, are strictly related to Fourier Transform and Weyl-Heisenberg algebra.
We discuss here the addition of a new discrete variable that measures the degree of the Hermite functions and allows to obtain the projective algebra $io(2)$.
A Rigged Hilbert space is found and a new discrete basis in $\R$ obtained.
The operators $\{{\cal O}[\R]\}$ defined on $\R$ are shown to belong to the Universal Enveloping Algebra UEA[$io(2)$] allowing, in this way, their algebraic discussion.
Introducing in the half-line a Fourier-like Transform, the procedure is extended to\, $\R^+$\,   and can be easily generalized to $\R^n$ and to spherical reference systems.
\end{abstract}

\section{INTRODUCTION}

The starting point of Quantum Mechanics in the line\, $\R$\,
(and its wave functions counterpart $L^2(\R)$)
are the
Weyl-Heisenberg algebra (WHA), $\langle X, P, \I\rangle$ \cite{Co} that defines
position and momentum  (or time and frequency) and the
Fourier Transform [FT] \cite{Fo} that relates them each other.
This approach is easily extended to Signal Processing both in Informatics
and Optics also if, there, the operatorial structure is scarcely  taken in consideration.

We consider here a more predictive approach
based on the Hermite functions (HF) \cite{NIST},
introduced as transition matrices between discrete and continuous bases
in a quantum space~\cite{CeOl}.

The central point is to add to the WHA a new operator\, $N$\, that reads the degree
of the HF. A Rigged Hilbert Space (RHS) \cite{Bo} is found and a projective algebra  $io(2)$ \cite{Ha,Ba} constructed where
$N$\, defines a new discrete basis in\, $\R$\, and $L^2(\R)$. Moreover the space of the operators
$\{{\cal O}[\R]\}$ (and\, $\{{\cal O}[L^2(\R)]\}$)
is isomorphic to the  Universal Enveloping Algebra of $io(2)$ (UEA[$io(2)$]).

Three bases are indeed obtained, two continuous, $\{|x\rangle\}$ and $\{|p\rangle\}$
related to the position $X$ and momentum $P$ of WHA and a new one, discrete,
$\{|n\rangle\}$ connected to the HF:
as in a Hilbert space (HS) all bases have the same cardinality, the mathematical structure
is a RHS, the true space of
Quantum Mechanics. A state in Quantum Mechanics was indeed originally defined
as a  ray in  a RHS,
an intricate concept involving a Gelfand triple\;  $\phi \subseteq {\cal H} \subseteq \phi'$ :
${\cal H}$ is a HS,\;  $\phi$ (dense subset of  ${\cal H}$) is the  $``ket"$ space
and\,  $\phi'$ (dual of  $\phi$)\, is the  $``bra"$ space. However quickly RHS has been considered an unnecessary complication
as all results can be found (after elaborated limits on functions with compact support) in the HS obtained representing, by means of the axiom of choice, each entire ray by a vector of norm one and phase zero.

We attempt to show here that RHS are not so complicated to justify its elimination from all recent textbook and can be more powerful that the standard HS because they allow an operatorial description  and an inclusion inside the same algebra
of variables with different  cardinality.
Hoping that it could  convince more than a general discussion, that can be found in \cite{Bo}, we discuss here the 1-dimensional case.

The projective algebra $io(2)\equiv \langle X, P, N, \I\rangle$ above introduced
is isomorphic to the so called algebra of harmonic
oscillator $\langle H, a, a^\dagger, \I\rangle$ \cite{Ha,Ba}, usually described inside the UEA[WHA].
As the $\infty$-dimensional UEA[$io(2)$] is a complex object,
a set of relevant substructures
emerges in\, $\R$\, and $\{{\cal O}[\R]\}$\; as well as in $L^2(\R)$ and $\{{\cal O}[L^2(\R)]\}$.
In particular, as $\infty$-many copies of $io(2)$ are contained into UEA[$io(2)$],
$\infty$-many operators that close $io(2)$ are found.

It should be noted that in $L^2(\R)$ (but not in $\R$, $\{{\cal O}[\R]\}$ and
$\{{\cal O}[L^2(\R)]\}$) the subspaces  structure derived from the algebra could
be obtained also from the Fractional Fourier Transform ([FrFT]) \cite{Oz}.
Indeed the eigenvectors of the [FrFT] are, as the ones of  [FT], the Hermite functions and
for each $k \in \N$ a [FrFT] can be constructed such that its eigenvalues are the
$k$th roots of unity, each one corresponding to one subspace.

To exhibit that the approach in not peculiar of the line\, $\R$, the discussion is repeated on the half line\, $\R^+$.
For this purpose, starting from the HF and the relations between HF and generalized Laguerre functions, two Fourier-like
Transforms [T$^\pm$] are constructed  with appropriate eigenvectors
on the half-line.

The extension to   $\R^n$ and to spherical coordinates will be discussed elsewhere.

\section{THE\, LINE\; $\R$}

To describe the line we start from the unitary irreducible representations of the translation group\; $T^1$
\[
P\, |p\rangle\; =\; p\; |p\rangle,\qquad \qquad U^p(x)\; |p\rangle\; =\; e^{-{\bf i} p x}\; |p\rangle .
\]
The regular representation\; $\{|p\rangle\}$\;  \; ($-\infty \,<\, p \,<\, \infty$) is such that
\[
\langle\, p\,|\,p'\,\rangle\; = \sqrt{2 \pi}\; \delta(p-p'),\qquad\qquad
\quad \frac{1}{\sqrt{2 \pi}}\;\int_{-\infty}^{+\infty} |p\rangle\, dp\, \langle p|\, =\, {\mathbb I},
\]
and the conjugate basis\, $\{|x\rangle\}$, defined by the operator $X$, is obtained
using the [FT]
\[
|x\rangle:=\left[\frac{1}{\sqrt{2 \pi}}
\int_{-\infty}^{+\infty}\, dp\; e^{-{\bf i} p x}\right]|p\rangle,
\qquad
\langle\, x\,|\,x'\,\rangle\; = \sqrt{2 \pi}\; \delta(x - x'),\qquad
\; \frac{1}{\sqrt{2 \pi}}\int_{-\infty}^{+\infty}\, |x\rangle\, dx\, \langle x|\,  =\, {\mathbb I} .
\]
The operators $X$ and $P$  together with $\I$ close the WHA \cite{Co}. At this point, we move to a non standard approach related to the ray
representations \cite{Ha, Ba, CeTa} of the inhomogeneous orthogonal algebra $io(2)$. In addition to $X$ and $P$ (which spectra have the cardinality  of the
continuum\;  $\aleph_1$)
an operator $N$ (with spectrum of cardinality $\aleph_0$), related to the index
$n$ of the Hermite functions (HF)
\[
\psi_n(x):=\; \frac {e^{-x^2/2}}{\sqrt{2^n n! \sqrt{\pi}}}\; H_n(x) ,
\]
(where\; $H_n(x)$\; are the Hermite polynomials) is introduced.
As the Hermite functions are  a basis of the space of  square
integrable functions on the line\; $L^2((-\infty, \infty))\, \equiv\, L^2(\R)$ \cite{Fo}:
\[
\int_{-\infty}^{\infty} \psi_n(x)\; \psi_{n'}(x)\; dx = \delta_{n,n'},\qquad\qquad
\sum_{n=0}^\infty\,  \psi_n(x)\;\, \psi_n(x')\, =\; \delta(x-x') ,
\]
the basic idea is now to introduce the vectors\; $\{|n\rangle\}$
\begin{equation}\label{defn}
|n\rangle\; :=\; (2 \pi)^{-1/4}\,\int_{-\infty}^{\infty}\, dx\; \psi_n(x)\; |x\rangle ,
\qquad  n \in \N\; ,
\end{equation}
that, by inspection, are an orthonormal and complete set in\, $\R$
\[
\langle\, n\,|\,n'\,\rangle
= \delta_{n\,n'},\qquad\qquad
\sum_{n=0}^{\infty}\; |n\rangle\, \langle n|\;  =\; \mathbb I  .
\]
The set $\{|n\rangle\}$\; is thus a discrete basis in the real line\; $\R$.

Relations among the three bases are easily established, as\; $\{\psi_n(x)\}$\, are
eigenvectors of [FT],
\[\label{FT}
[{\rm FT}]\; \psi_n(x)\;=\left[\frac{1}{\sqrt{2 \pi}}
\int_{-\infty}^{\infty} dx\; e^{\,{\bf i} p x}\right]\psi_n(x)\; \equiv\; \frac{1}{\sqrt{2 \pi}}\int_{-\infty}^{\infty} dx\;
e^{\,{\bf i} p x}\;\psi_n(x)\; =\; {\bf i}^n\; \psi_n(p),
\]
\[\begin{array}{llllll}
|x \rangle &=&(2 \pi)^{1/4}\; \sum_{n=0}^\infty \, \psi_n(x)\; |\,n\rangle , \qquad\qquad
&|p\,\rangle &=&\left[\frac{1}{\sqrt{2 \pi}}
\int_{-\infty}^{\infty}\, dx\,
e^{\, {\bf i} p x}\right]\, |x\rangle ,
\\[0.5cm]
 |n\rangle &=& {\bf i}^n\; (2 \pi)^{-1/4}  \,\int_{-\infty}^{\infty}\,
dp\; \psi_n(p) \; |p\rangle ,\qquad\qquad
&|p\, \rangle &=& (2 \pi)^{1/4}\; \sum_{n=0}^\infty  \,{\bf i}^n\; \psi_n(p)\; |n \rangle .
\end{array}\]

For an arbitrary vector\; $|f\rangle$ $\in$ $\R$ we thus have
\[
|f\rangle\;\;  =\;\;\frac{1}{\sqrt{2\pi}} \int_{-\infty}^{+\infty}\, dx\;
f(x)\; |x\rangle \quad  = \quad
\frac{1}{\sqrt{2\pi}} \int_{-\infty}^{+\infty}\, dp\; f(p)\; |p\rangle \quad  =\quad
\sum_{n=0}^\infty \, f_n\; |n\rangle\, ,
\]
\[\begin{array}{l}
f(x)\, :=\; \langle x|f\rangle\;  =\;(2 \pi)^{1/4} \sum_{n=0}^\infty\, \psi_n(x)\;f_n , \qquad
f(p)\, :=\; \langle p|f\rangle\;  =\;(2 \pi)^{1/4} \sum_{n=0}^\infty\,(-{\bf i})^n\, \psi_n(p)\;f_n ,
\\[0.5cm]
f_n\, :=\; \langle n  | f\rangle\;  =\;\; (2 \pi)^{-1/4} \int_{-\infty}^{+\infty} \,dx\;\psi_n(x)\;
f(x)
\;\;  =\;\;\;  {\bf i}^n(2 \pi)^{-1/4}\int_{-\infty}^{+\infty} \,dp\;  \psi_n(p)\, f(p),
\end{array}\]
and the wave functions\, $f(x)\,, f(p)$ and the sequence\, $\{f_n\}$\; describe
$|f\rangle$ in the three bases.

All seems trivial, but\;  $\{|n\rangle\}$\;  has the cardinality of the natural numbers\;  $\aleph_0$ and, as  all bases in a Hilbert space have the same cardinality,
the structure we have constructed (the quantum space on the line $\R$) is not
an Hilbert space but a Rigged Hilbert space.

In this way, by means of [FT] and HF, we went back to the foundations  of Quantum Mechanics, where a physical state was defined  as a ray in a Hilbert space and not as a vector in HS as we are used, considering the RHS as a unnecessary complication.
However the problem is apparent also in the simple case of the harmonic oscillator: the algebraic description consider\;  $H\, (\equiv N + 1/2)$\,,\,  $a$,\, $a^\dagger$\, and\, $\I$. But
$X = \frac{1}{\sqrt{2}} (a+a^\dagger)$\, and $P \equiv -{\bf i}\,
D_x = \frac{-{\bf i}}{\sqrt{2}}(a-a^\dagger)$\;  do not belong to
the algebra because they
are not operators in a HS.
In a RHS, instead, $X$ and $P$ are standard operators also if their spectra are continuous and we have no problems to consider them as generators of a Lie algebra together
with the number operator $N$ \cite{CeOl, Bo}.
This allows us to extend the\;  Lie algebra\; to the differential structure and, in general, to include all operators inside the involved Universal Enveloping Algebra.

To describe the structure of the operators on the line $\R$
we thus begin introducing in the $L^2(\R)$\, the operators
$X ,\; D_x (= {\bf i} P), N$ and  ${\mathbb I}$
\[
 X \psi_n(x) :=  x\, \psi_n(x),\quad  D_x \psi_n(x) := \psi_n'(x),\quad
N \psi_n(x) :=  n\, \psi_n(x),\quad  {\mathbb I}\, \psi_n(x) := \psi_n(x) .
\]
The recurrence relations of Hermite polynomials
\[
H'_n(x)\, =\, 2\, n\, H_{n-1}(x),\qquad  H'_n(x) - 2\, x\, H_n(x)\, =\, H_{n+1}(x),
\]
can be rewritten as
\[
a\;\, \psi_n(x)\, =\, \sqrt{n}\;\, \psi_{n-1}(x) ,\;\qquad
a^\dagger\;\, \psi_n(x)\, =\, \sqrt{n + 1}\;\, \psi_{n+1} (x),
\]
where the operators $a$ and $a^\dagger$ are defined in terms of the
operators $X$ and $P$
\[
a:=\, \frac{1}{\sqrt{2}}\, \left(X+{\bf i} P\right),\qquad
a^\dagger :=\, \frac{1}{\sqrt{2}}\, \left(X-{\bf i} P\right).
\]
The algebra contains indeed
the rising and lowering operators on the HF \cite{Ban, Lo}
 \begin{equation}\label{ho}
[N, a^\dagger] = a^\dagger ,\qquad[N, a] = -a , \qquad [a,a^\dagger]= \mathbb I ,
\qquad[ \mathbb I ,\bullet] = 0 ,
\end{equation}
and is isomorphic to the projective algebra $io(2)$ \cite{Ha, Ba}:
\begin{equation}\label{io2}
[N,X] = -{\bf i} P,\qquad [N,P] = {\bf i} X,\qquad [X,P] = {\bf i}\, \I ,\qquad
[ \mathbb I ,\bullet] = 0 .
\end{equation}

Because of eq. (\ref{defn}) the RHS  $\R$ and $L^2(\R)$ are isomorphic and support  a representation  with zero value of the Casimir operator
of $iso(2)$ (\ref{io2}) (or equivalently  (\ref{ho})), as discussed in \cite{CeTa},
\begin{equation}\label{C}
{\cal C}\, \equiv\, (X^ 2 - D_x^2)/2 - N - 1/2\, =\,  \{a, a^\dagger \}/2 -(N + 1/2)\,\I =0 .
\end{equation}
So, in the RHS of the line $\R$, we have
\[
{\cal C}\, |n\rangle\; =\; \left[ \{a, a^\dagger \}/2 -(N + 1/2)\,\I\right] |n\rangle\; =\;0 ,
\]
that in $L^2(\R)$ (the RHS of square integrable functions defined on the line) can also be written as
\[
{\cal C}\; \psi_n(x)\; =\; \left[(X^ 2 - D_x^2)/2 - N - 1/2 \right ]\; \psi_n(x)\; =\; 0 .
\]
Eq. (\ref{C}) is indeed the operatorial identity that defines the RHS $\R$ and $L^2(\R)$
\begin{equation}\label{id}
N\; \equiv\; \left(X^ 2 - D_x^2 - 1\right)/2\; \equiv\; \{a,a^\dagger\}/2 -1/2 ,
\end{equation}
by inspection equivalent to the Hermite differential equation
\[
H''_n(x) - 2\, x\, H'_n(x) + 2\, n\, H_n (x) = 0 .
\]

The representation is irreducible, so that,
on both spaces $L^2(\R)$\,and \;  $\R$,
all operators of the UEA[$io(2)$] are defined and
an isomorphism exists between the UEA[$io(2)$] and the space
of the operators\;  $\{{\cal O}[L^2(\R)]\}$\;  and $\{{\cal O}[\R]\}$
\[\label{iso}
 \{{\cal O}[L^2(\R)]\}\, \equiv\, UEA[io(2)] \,\equiv\, \{{\cal O}[\R]\},
\]
i.e. each operator\;\; ${\cal O}$\;\; can be written
\[
{\cal O}\; = \sum c_{\alpha\,\beta\,\gamma}\; X^\alpha\, {D_x}^{\beta}\, N^\gamma
=\; \sum d_{\alpha\,\beta\,\gamma}\; {a^\dagger}^\alpha\, N^\beta\, a^\gamma .
\]

From the analytical point of view, an ordered monomial\;\;
$X^\alpha\, {D_x}^{\,\beta}\, N^\gamma  \in {\rm UEA}[io(2)]$
is an order\, $\beta$\, differential operator but, because of the
operatorial identity (\ref{id}), we have
\[
D_x^2\; \equiv\; X^ 2 - 2 N - 1 ,
\]
and any operator in $\{{\cal O}[L^2(\R)]\}$ and in $\{{\cal O}(\R)\}$ can be written
\[
{\cal O}\, =\, f_0(X)\, g_0(N)\; +\; f_1(X)\; D_x\; g_1(N),
\]
that on the basis vector\, $\psi_n(x)$\; becomes
\[
{\cal O}\; \psi_n(x)\; = \; f_0(x)\, g_0(n)\, \psi_n(x)\; +\; f_1(x)\, g_1(n)\; {\psi'}_n(x) .
\]

The\, UEA[$io(2)$]\,
 is a rich structure. In particular it  contains\, $\infty$-many\,
 $io(2)$ algebras that allow to construct an intricate structure of subspaces. In particular
for any $k \in \N$ we have
\[
{\rm UEA}[io(2)]\;\; \supset \;\;\;\oplus_{r=0}^{k-1}\; io_{k,r}(2),
\]
where\, $k$ and $r$\, are parameters that define each copy $io_{k,r}(2)$ of the algebra $io(2)$.
Indeed, starting from\, $n$\, and\, $k$\;, we can define two other integers
$q =$ Quotient$[n,k]$\;\, and\;\, $r =$ Mod$[n,k]$\; so that\;  $n=k\, q+r$ and the
 operators $Q$ and $R$  are, of course, diagonal on\; $\{|k\; q+r\rangle\}$
\[
Q\, | k q+r\rangle =  q\, | k q+r\rangle \qquad R\, | k q+r\rangle = r\, | k q+r\rangle .
\]
By inspection, the operators\; $A_{k,r}^\dagger, A_{k,r} \in {\rm UEA}[io(2)]$
\begin{equation}\label{Akq}
A_{k,r}^\dagger := (a^\dagger)^k  \frac{\sqrt{N+k-r}}{\sqrt{k \prod_{j=1}^{k}(N+j)}},\qquad
A_{k,r} := \frac{\sqrt{N+k-r}}{\sqrt{k \prod_{j=1}^{k}(N+j)}} (a)^k
\end{equation}
are defined on the whole set $\{|k\, q+r\rangle\}$ and give
\[
A_{k,r}^\dagger\, |k\, q+r\rangle =  \sqrt{q+1}\; |k(q+1)+r\rangle ,\qquad
A_{k,r}\, |k\, q+r\rangle = \sqrt{q}\; |k(q-1)+r\rangle .
\]
Each couple, $k$ and $0\leq r<k$, gives us a irreducible representation with\;
${\cal C} = 0$\; of\; $io(2)$ that we denote  $io_{k,r}(2)$
\[
[Q,\, A_{k,r}^\dagger]\,=\, +\, A_{k,r}^\dagger,\qquad\; [Q, A_{k,r}]\,=\, -\, A_{k,r},\qquad\;
[ A_{k,r},\, A_{k,r}^\dagger] = \mathbb I ,\qquad\;[ \mathbb I ,\bullet] = 0 .
\]
In particular, for $k=4$ and $0\leq r<4$, we can define
\[
\R_{4,r} := \{|4\,q+r\rangle\}, \qquad \qquad  L_{4,r}^2(\R) := \{\psi_{4\, q +r}(x)\},
\qquad\qquad\quad  (q=0,1,2,\dots)
\]
and\, $L^2(\R)$\;  and \;$\R$\;  can be split in four subspaces
each one representation of one of  $io_{4,r}(2)$
\begin{eqnarray}\label{h4}
L^2(\R)\; &=&\; \oplus_{r=0}^3 \;\, L_{4,r}^2(\R) =\; \oplus_{r=0}^3
 \;\, \{\psi_{4 \,q+r}(x)\},
\\[0.4cm]
\R\; &=&\; \oplus_{r=0}^3 \;\,\R_{4,r}\;\;\quad =\; \oplus_{r=0}^3
 \;\, \{|4 \,q+r\rangle\} .\nonumber
\end{eqnarray}
For each $0\leq r<4$  the operators
$\{{\cal O}[L_{4,r}^2(\R)]\}$\, and
$\{{\cal O}[\R_{4,r}]\}$ belong to UEA$[io_{4,r}(2)]$
\[
 \{{\cal O}[L^2(\R_{4,r})]\}\, \equiv\, {\rm UEA}[io_{4,r}(2)] \,\equiv\, \{{\cal O}[\R_{4,r}]\}\; .
\]

These results are general since eqs. (\ref{Akq}) allow to write  for all\, $k\in \N$,
$0\leq r<k$\, and\, $q =0,1,2,\dots$
\[\begin{array}{cllcll}
L_{k,r}^2(\R) &:=& \{\psi_{k \,q+r}(x)\}, \quad
&\R_{k,r} &:=& \{|k \,q+r\rangle\},
\\[0.5cm]
L^2(\R) &=& \oplus_{r=0}^{k-1} \;\, L_{k,r}^2(\R) =\, \oplus_{r=0}^{k-1}\;
 \, \{\psi_{k \,q+r}(x)\} ,\quad
 &\R & =& \oplus_{r=0}^{k-1} \; \R_{k,r}  = \oplus_{r=0}^{k-1}
 \; \{|k\;q+r\rangle\}.
 \end{array}\]
 Hence
\[
 \{{\cal O}[L^2(\R_{k,r})]\}\, \equiv\, {\rm UEA}[io_{k,r}(2)] \,\equiv\, \{{\cal O}[\R_{k,r}]\} .
\]

The eq.  (\ref{h4}) is well-known from the  Fourier Transform (\ref{FT}) that splits $L^2(\R)$
in four subspaces each one corresponding to one eigenvalue of [FT] (but, of course,
does not consider the operators). Its generalization to all $k\in \N$, that allows to divide $L^2(\R)$ in $k$ subspaces
(again disregarding the operators), can be obtained from the Fractional Fourier Transform ([FrFT])~\cite{Oz}:
\[
[{\rm FrFT}]_\alpha\; \psi_n(x) =\, e^{ {\bf i} \alpha n}\, \psi_n(x'),
\qquad\qquad  \alpha \in \C,
\]
by specializing $\alpha$ to  $\alpha= 2 \pi/k$
\begin{equation}\label{FrFT}
[{\rm FrFT}]_{2 \pi/k}\;\, \psi_n(x) = e^{{\bf i} \frac{2 \pi n}{k}}\; \psi_n(x'),
\end{equation}
\[
[{\rm FrFT}]_{2 \pi/k}\;\, f(x)\, :=\, \frac{1}{\sqrt{2 \pi}} \int_{-\infty}^{\infty}\, dx\; e^{{\it i} x x'}\,
e^{2 \pi {\bf i}(1/k-1/4)  N}\; f(x)\;=\,
[{\rm FT}]\left[ e^{2 \pi {\bf i}(1/k-1/4)  N}\; \right]f(x),
\]
where\;  $N$\; --diagonal on each\, $\psi_n(x)$--\; on the whole space is rewritten as\;
$N\; \equiv\; ({X}^ 2 - D_{x}^2 - 1)/2$.

As an example, for\, $k=3$\, , eq. (\ref{FrFT}) gives
\[
[{\rm FrFT}]_{2 \pi/3}\; \psi_n(x) =\, e^{{\it i}\frac{2 \pi n}{3}}\, \psi_n(x'),
\]
that exhibits three subspaces of $L^2(\R)$
\[
L^2(\R) =\, L^2_{3,0}(\R)\, \oplus\, L^2_{3,1}(\R)\, \oplus\, L^2_{3,2}(\R)\, =\,
\{\psi_{3\, q}\}\oplus\, \{\psi_{3\, q+1}\}\oplus \{\psi_{3\, q+2}\} .
\]

\section{THE\, HALF-LINE\; $\R^+$}

The results obtained on the line\, $\R$\, can be rewritten in\, $\R^+$, the vector space
defined by the operator\;  $Y$\; ,
with basis\; $\{|y\rangle\}$\; with $y$ into the open set $(0,+\infty)$:
\[
Y\, |y \rangle\; =\; |y\rangle\, y ,\qquad\qquad
\langle \,y\,|\,y'\,\rangle\, =\, \delta(y-y') , \qquad\qquad\;
 \int_{0}^{\infty}\,  |y\rangle \,dy\, \langle y|\, =\, \mathbb I  \,.
\]

The generalized Laguerre polynomials \;  $L_n^\alpha(y)$\,\, ($n \in \N$) play now the role of the Hermite polynomials.
For each value of $\alpha$ an alternative discrete basis $\{|n\rangle\}$ is obtained.
From $L_n^\alpha$(y)\,  the normalized generalized
Laguerre functions\; ${M_n^\alpha}(y)$\,are defined by
\[
{M_n^\alpha}(y) :=\; \sqrt{\frac{\Gamma(n+1)}{\Gamma(n+\alpha+1)}}\;\;\;
y^{\alpha/2}\;\; e^{-y/2}\;\,  {L_n^\alpha}(y),
\]
that satisfy
\[
\displaystyle
 \int_{0}^{\infty} M_{n}^{\alpha}(y)\, \, M_{m}^{\alpha}(y) \;dy\;=\; \delta_{n m}\, ,\qquad\qquad
\displaystyle \sum_{n=0}^{\infty}  \;M_{n}^{\alpha}(y)\,\, M_{n}^{\alpha}(y')\;
=\;
\delta(y-y').
\]
Hence, like the case of  $\{\psi_n(x)\}$\, in\, $L^2(\R)$\,, for any  fixed $\alpha$ the set
$\{M^{\alpha}_n(x)\}$\, is a basis in\,
$L^2(\R^+)$~\cite{Fo}.
Again like in (\ref{defn}) we define, for $n\in \N$ and fixed $\alpha$, the vector\;  $|n\rangle$
\[
|n\rangle\,:=\;\; \int_{0}^{\infty}  dy\; M_n^\alpha(y)\; |y\rangle \, ,
\]
and we have
 \[
\langle n| n'\rangle = \delta_{n\,m}\, ,\qquad\qquad
 \sum_{n=0}^{\infty}\; |n\rangle \langle n|\, =\,   \,\mathbb I ,
\]
i.e. the set \,$\{|n\rangle\}_{n\in \N}$\, is an orthonormal basis
of the half-line $\R^+$ and the generalized Laguerre functions
$M_n^\alpha(y)$ are the transformations matrices that relate the two
bases $\{|y\rangle\}$ and $\{|n\rangle\}$
\[
M_{n}^{\alpha}(y)\, =\;\, \langle \,y\,|\,n\,\rangle\;\; =\;\,  \langle \,n\, |\,y\,\rangle\, .
\]
Like in the line, an arbitrary vector\; $|f\rangle$ $\in$
 $\R^+$
can be written
\[
|f\rangle = \int_{0}^{\infty}\,  dy\, f(y)\, |y\rangle \; \quad = \quad
\sum_{n=0}^\infty \;\, |n\rangle\, f_n \, ,
\]
\[
f(y):=\, \langle y|f\rangle\, =\, \sum_{n=0}^\infty\, M_n^\alpha(y)\; f_n ,\qquad\;
f_n:=\, \langle n  | f\rangle\, = \int_{0}^{\infty} \,dy\;M_n^\alpha(y)\; f(y),
\]
so that the wave function\, $f(y)$\, and the sequence\, $\{f_n\}$\;
describe the vector $|f\rangle$ in the two bases.

Since a  discrete basis $\{|n\rangle\}$ and a continuous one $\{|y\rangle\}$ have been
found also the half-line is a RHS.


We introduce now the algebra.
On the $L_n^\alpha(y)$ the rising and lowering operators are
\[
\begin{array}{rll}\displaystyle
y\, L_n^\alpha(y)' +(n+\alpha+1-y)\, L_n^\alpha(y) &=& (n+1)\, L_{n+1}^\alpha(y),
\\[0.5cm] \displaystyle
-y\, L_n^\alpha(y)' +n\, L_n^\alpha(y) &=& (n+\alpha)\, L_{n-1}^\alpha(y)\; ,
\end{array}
\]
that, defining the operators $Y$, $D_y$, $\N$ and $\I$
\[
Y\, M_n^\alpha(y) := y\, M_n^\alpha(y),
\quad D_y\, M_n^\alpha(y):= M_n^\alpha(y)' ,\quad
N\,  M_n^\alpha(y) :=  n\,  M_n^\alpha(y),\quad
 {\mathbb I}\,  M_n^\alpha(y) := M_n^\alpha(y),
\]
can be rewritten as
\begin{equation} \begin{array}{lllll}\label{jpm}\displaystyle
J_+\; M_n^\alpha(y)&:=& \left(
Y\, D_y +N+1 +\frac{\alpha-Y}{2}\,
\right)
M_n^\alpha(y) &=& \sqrt{(n+1)(n+\alpha+1)}\; M_{n+1}^\alpha(y),
\\[0.5cm] \displaystyle
J_-\; M_n^\alpha(y)&:=& \left( -Y\, D_y +N +\frac{\alpha-Y}{2}
\right) M_n^\alpha(y) &=&
\sqrt{n(n+\alpha)}\; M_{n-1}^\alpha(y)\,  .
\end{array}\end{equation}

By inspection we see that \,  $J_\pm = {J_\mp}^\dagger$.
Note that, unlike the full line, the operator $N$ is needed to write
the rising and lowering operators.
To complete eqs. (\ref{jpm}), we introduce now\; $J_3:= N+(\alpha+1)/2$ acting as
\begin{equation}\label{j3}
J_3\; M_n^\alpha(y) = \left[N+(\alpha+1)/2\right]\, M_n^\alpha(y)\, =\,
(n+(\alpha+1)/2)\; M_n^\alpha(y),
\end{equation}
that, together with $J_\pm$ close the $su(1,1)$ algebra
\[
[J_3, J_\pm] = \pm J_\pm , \qquad [J_+, J_-] =  - 2\, J_3\, .
\]
From eqs. (\ref{jpm}) and (\ref{j3}) we see that  the $\{M_n^\alpha(y)\}$\; defines an irreducible representation  of
this\;  $su(1,1)$\; algebra.
Furthermore, as $\{|n\rangle\}$\, is isomorphic to\; $\{M_n^\alpha(y)\}$\,, we have
\[
\begin{array}{l}\nonumber
J_+\, |n\rangle\, =\; \sqrt{(n+1)(n+\alpha+1)}\; |n+1\rangle ,\\[0.4cm]
J_3\,\, |n\rangle\, =\; \left(n+(\alpha+1)/2\right) \; |n\rangle ,\\[0.4cm]
J_-\, |n\rangle\, =\, \sqrt{n(n+\alpha)}\;\, |n-1\rangle .
\end{array}
\]
The Casimir operator ${\cal C}$ \,is
\[
{\cal C}= \left(J_3^{\,2}-\frac 12 \{J_+,J_-\}\right)\, =\,
 \frac{\alpha^2-1}{4}\;\,
\]
that, remembering eqs. (\ref{jpm}) and (\ref{j3}), can be written as
\begin{equation}\label{de}
\left[Y\, D^{~2}_y + D_y +N +\frac{\alpha+1}{2} -\frac{\alpha^2}{2 Y}
- \frac{Y}{4}\right] M_n^\alpha(y)\; =0,
\end{equation}
which is equivalent to the associated Laguerre equation
\[
y\; L_n^\alpha(y)''+ (\alpha+1-y)\, L_n^\alpha(y)'
+n\; L^\alpha_n(y) = 0\,.
\]
We have thus on the half-line the operatorial identity
\begin{equation}\label{id2}
N \equiv\;
 -Y\, D^{~2}_y - D_y -\frac{\alpha+1}{2} +\frac{\alpha^2}{4 Y} + \frac{Y}{4} .
\end{equation}

Also in the half line, the representation is irreducible so that
the operators acting on  the $L^2(\R^+)$\, and\, $\R^+$\, belong to the\, UEA$[ su(1,1)]$
i.e.\, they can be written as
\[
{\cal O}\; =\; \sum c_{\alpha\,\beta\,\gamma}\; {J_+}^\alpha\, {J_3}^\beta\, {J_-}^\gamma \;=\;
\sum d_{\alpha\,\beta\,\gamma}\; Y^\alpha\, {D_y}^{\beta}\, N^\gamma ,
\]
and the space of the operators\;  $\{{\cal O}[L^2(\R^+)]\}$\;  and $\{{\cal O}[\R^+]\}$)
are isomorphic  to the UEA$[su(1,1)]$
\[
 \{{\cal O}[L^2(\R^+)]\}\, \equiv\, {\rm UEA}[su(1,1)] \,\equiv\, \{{\cal O}[\R^+]\} .
\]

The monomials \, $ {J_+}^\alpha\, {J_3}^{\beta}\, {J_-}^\gamma  \in {\rm UEA}[su(1,1)]$\;\; look to be differential operators
of the order\;\,  $\alpha + \gamma$\;\, but the identity (\ref{id2}) can also be read
\[
D^{~2}_y \; \equiv \;-\,\frac{1}{Y} \left(D_y +N +\frac{\alpha +1}{2} -\frac{\alpha^2}{2 Y} -\frac{Y}{4}\right),
\]
so that all  the off-diagonal operators are at most first
order differential operators and all the diagonal ones are  equivalent
to eq. (\ref{de}).

Also the UEA[$su(1,1)$]\,  contains, in analogy with
the UEA[$io(2)$]\,,  $\infty$-many $su(1,1)$, \, denoted by $su_{k,r}(1,1)$\,.
We can define $L_{k,r}^2(\R^+) = \{M_{k\,q+r}(\R^+)\}$\;, \,$\R_{k,r}^+ = \{|k\,q+r\rangle\}$
and for each\, $k \in \N$\,  the spaces $L^2(\R^+)$\, and\, $\R^+$\, are direct sums of $r$ depending subspaces
\[
L^2(\R^+) = \oplus_{r=0}^{k-1}\; L_{k,r}(\R^+)\qquad\quad
\R^+ = \oplus_{r=0}^{k-1}\;  R_{k,r}^+ .
\]
The same subspace structure is found in the spaces of operators $\{{\cal O}[L^2(\R^+)]\}$ and $\{{\cal O}(\R^+)\}$, i.e. we have
\[
 \{{\cal O}[L^2(\R_{k,r}^+)]\}\, \equiv\, {\rm UEA}[su_{k,r}(1,1)] \,\equiv\, \{{\cal O}[\R_{k,r}^+]\} .
\]

\section{FOURIER-LIKE\, TRANSFORMS\, ON\, $\R^+$}

To find the third basis, conjugate  of\, $\{|y\rangle\}$, we should need now something that
plays in $\R^+$ the role of the [FT],  i.e. an Integral Transform  that has
as eigenvectors the $\{M^{\alpha}_n\}$.
We have not been able to obtain it for generic $\alpha$ but only for $\alpha = \pm 1/2$.
In these cases, remembering the well known relations \cite{NIST}
\[
\psi_{2 n}(x) = (-1)^n (x^2)^{1/4} M_n^{-1/2}(x^2),\qquad\;\;
\psi_{2 n+1}(x) = (-1)^n x (x^2)^{-1/4}\, M_n^{+1/2}(x^2),
\]
and the fact that Hermite functions are eigenstates of [FT],
we can find two\, new transforms\;  [T$^\pm$]
\[\begin{array}{l}
\left[{\rm T}^+\right]\, f(y)\, :=\;
\displaystyle\left[\frac{(-1)^n}{\sqrt{2 \pi}} \int_0^\infty \; dy\,
\frac{\sin (\sqrt{y y'})}{(y y')^{1/4}}\right]\;\; f(y),
\\[0.5cm]
\left[T^-\right]\, f(y)\, :=\;
\displaystyle \left[\frac{(-1)^n}{\sqrt{2 \pi}} \int_0^\infty \;dy\;
\frac{\cos (\sqrt{y y'})}{(y y')^{1/4}}\right]\;\, f(y)
\end{array}\]
that have $M_n^{\pm 1/2}(y)$\; as eigenvector and\, $(-1)^n$ as eigenvalues
\[
\left[T^\pm\right]\; M_n^{\pm1/2}(y) =\; (-1)^n\; M_n^{\pm1/2}(y') .
\]

In conclusion, two alternative conjugate basis of $\{|y\rangle\}$ have been found
\[
|q\rangle^+  =\; \frac{(-1)^n}{\sqrt{2 \pi}} \int_0^\infty dy\;
\frac{\sin (\sqrt{q y})}{(q y)^{1/4}}\;\,|y\rangle , \qquad\quad
|q\rangle^-  =\; \frac{(-1)^n}{\sqrt{2 \pi}} \int_0^\infty dy\;
\frac{\cos (\sqrt{q y})}{(q y)^{1/4}}\;\,|y\rangle
\]
and two different substructures obtained
\[
L^2(\R^+) = L_0^2(\R^+)^\pm\; \oplus\; L_1^2(\R^+)^\pm ,
\]
where
\[
L_0^2(\R^+)^\pm \; =\; \{ M_{2n}^{\pm 1/2}\},\qquad \qquad  L_1^2(\R^+)^\pm \;=\; \{M_{2n+1}^{\pm 1/2}\}, \quad \forall n \in \N\, .
\]
This last result has been generalized introducing
two New Fractional Transforms\; [FrT$^\pm$]\; ,
\[\begin{array}{l}
\left[{\rm FrT}^\pm\right]_{2\pi/k}\, f(y)\, := \left[{\rm T}^\pm\right]\left[ \;e^{2\pi {\bf i}(1/k-1/2)N}\right]\; f(y)
\\[0.4cm]
[{\rm FrT}^\pm]_{2 \pi/k} \;\, M_n^{\pm 1/2}(y)
 = e^{\,{\bf i}\, 2 \pi n/k}\; M_n^{\pm 1/2}(y'),
\end{array}\]
where $N$ is diagonal on  $M_n^{\pm 1/2}(y)$
 and, in general,  can be obtained from the identity (\ref{id2}).

The hope that, for a generic $\alpha$, an Integral Transform $[T^\alpha]$, with
eigenvectors $\{M_n^\alpha\}$ could be found (allowing $\infty$-many
equivalent bases $\alpha$-dependent on the same footing)  seems reasonable.
We are working on it.

\section{CONCLUSIONS}
Rigged Hilbert spaces are shown to be more predictive than Hilbert spaces in Quantum Physics and Signal Processing in Optics and Informatics  since operators of different cardinality can be considered together.

In Rigged Hilbert spaces continuous and discrete bases exist with special functions as transformation matrices between them.

In a RHS operators of different cardinality  can be together generators of a Lie algebra
and/or elements of its Universal Enveloping Algebra.

An elaborated algebraic structure inside both, the Quantum space and the space of operators defined on it, is found.
In particular, an infinite set of substructures emerges  both in the space of the states and in the space of operators acting on them.

An alternative definition of an Integral Transform, as an operator that has special functions as
eigenvectors,  has been introduced.
\newpage

\end{document}